# Attacking Automatic Video Analysis Algorithms: A Case Study of Google Cloud Video Intelligence API

Hossein Hosseini*, Baicen Xiao*, Andrew Clark** and Radha Poovendran*

*Department of Electrical Engineering, University of Washington, Seattle, WA
**Department of Electrical and Computer Engineering, Worcester Polytechnic Institute, Worcester, MA
hosseinh@uw.edu, bcxiao@uw.edu, aclark@wpi.edu, rp3@uw.edu

## ABSTRACT

Due to the growth of video data on Internet, automatic video analysis has gained a lot of attention from academia as well as companies such as Facebook, Twitter and Google. In this paper, we examine the robustness of video analysis algorithms in adversarial settings. Specifically, we propose targeted attacks on two fundamental classes of video analysis algorithms, namely video classification and shot detection. We show that an adversary can subtly manipulate a video in such a way that a human observer would perceive the content of the original video, but the video analysis algorithm will return the adversary's desired outputs.

We then apply the attacks on the recently released Google Cloud Video Intelligence API. The API takes a video file and returns the video labels (objects within the video), shot changes (scene changes within the video) and shot labels (description of video events over time). Through experiments, we show that the API generates video and shot labels by processing only the first frame of every second of the video. Hence, an adversary can deceive the API to output *only* her desired video and shot labels by periodically inserting an image into the video at the rate of one frame per second. We also show that the pattern of shot changes returned by the API can be mostly recovered by an algorithm that compares the histograms of consecutive frames. Based on our equivalent model, we develop a method for slightly modifying the video frames, in order to deceive the API into generating our desired pattern of shot changes. We perform extensive experiments with different videos and show that our attacks are consistently successful across videos with different characteristics. At the end, we propose introducing randomness to video analysis algorithms as a countermeasure to our attacks.

## 1 INTRODUCTION

Machine learning (ML) techniques have substantially advanced in the past decade and are having significant impacts on everyday lives. In recent years, a class of ML algorithms called deep neural networks have been successfully deployed for computer vision tasks, particularly recognizing objects in images, where new

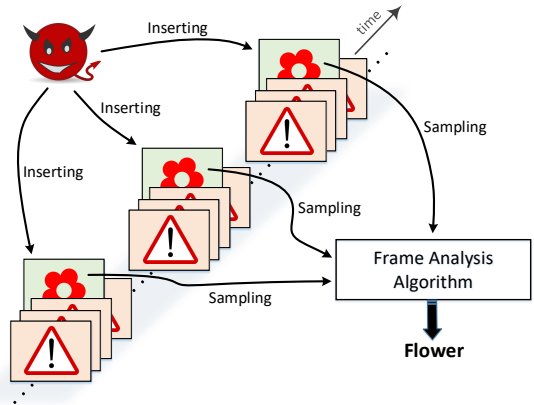

**Figure 1: Illustration of image insertion attack on video classification algorithms. The adversary modifies the video by placing her chosen image in the sampling locations of the algorithm. As a result, the generated video labels are only related to the inserted image.**

algorithms reported to achieve or even surpass the human performance [21, 28, 44].

With the growth of online media, surveillance and mobile cameras, the amount and size of video databases are increasing at an incredible pace. For example, in 2015, YouTube reported that over 400 hours of video are uploaded every minute to their servers [53]. Therefore, there is a need for automatic video analysis to handle the growing amount of video data. Automatic video analysis can enable searching the videos for a specific event, which is helpful in applications such as video surveillance or returning the search results on the web. It can be also used for prescanning user videos, for example in YouTube and Facebook platforms, where distribution of certain types of illegal content is not permitted.

Following the success of deep learning-based visual classification research, there has been a surge in research for video annotation [26, 27, 33, 57]. Internet companies such as Facebook [38] and Twitter [12] are also developing products for analyzing the videos on their platforms. Recently, Google has introduced the Cloud Video Intelligence API for video analysis [30]. A demonstration website [25] has been launched which allows anyone to test the API with videos stored in Google Cloud Storage. The API then quickly returns the *video labels* (key objects within the video), *shot changes* (scene changes within the video) and *shot labels* (description of every shot). By detecting the video shots and labeling them, the API can make videos searchable just as text documents. Thus, users can search for a particular event and get related videos along with the exact timings of the events within the videos.

ML models are typically designed and developed with the implicit assumption that they will be deployed in benign settings. However, many papers have pointed out their vulnerability in adversarial environments [4, 6, 7, 10]. Learning systems are increasingly applied in security-sensitive and critical systems, such as banking, medical diagnosis, and autonomous cars. This signifies the importance of studying their performance in adversarial settings.

In this paper, we examine the robustness of video analysis algorithms. Specifically, we propose targeted attacks against two fundamental classes of video analysis algorithms, namely video classification and shot detection. We then apply the attacks on Google Cloud Video Intelligence API. [1] We show that an adversary can subtly manipulate a video in such a way that a human observer would perceive the content of the original video, but the API will return the adversary's desired outputs. Such vulnerability will seriously undermine the performance of video analysis systems in real-world applications. For example, a search engine may wrongly suggest manipulated videos to users, a video filtering system can be bypassed by slightly modifying a video which has illegal content, or a video search algorithm may miss the important events in surveillance videos. Our findings further indicate the importance of designing ML systems to maintain their desired functionality in adversarial environments.

Our contributions are summarized in the following:

- We develop a model for state-of-the-art video classification and shot detection algorithms. We then formulate the adversary's objective function for mounting targeted attacks on black-box video analysis systems and discuss different approaches for video modification.
- We propose different methods for deceiving video classification and shot detection algorithms and demonstrate the effectiveness of our attacks on Google Cloud Video Intelligence API. In our experiments, we queried and tested the API with different videos, including our recorded and synthetically generated videos, videos downloaded from web, and the sample videos provided by API website. Selected videos vary in content, length, frame rate, quality and compression format.
- Through experiments, we show that the API's algorithm for generating video and shot labels processes only the first frame of every second of the video. Therefore, by inserting an image at the rate of one frame per second into the video, the API will *only* output video and shot labels that are related to the inserted image. The image insertion attack is illustrated in Figure 1.
- We also show that the pattern of shot changes returned by the API can be mostly recovered by finding the peaks in the vector of histogram changes of consecutive frames. Based on our equivalent model, we develop a method for slightly modifying the video frames, in order to deceive the API into generating our desired pattern of shot changes.

- We propose countermeasures against our attacks. We show that introducing randomness to video analysis algorithms improves their robustness, while maintaining the performance.

The rest of this paper is organized as follows. Section 2 provides a background on video data and describes the Cloud Video Intelligence API. Section 3 reviews video analysis methods. The threat model is given in Section 4. Sections 5 and 6 present our attacks on video classification and shot detection algorithms, respectively. Section 7 provides related works and Section 8 concludes the paper.

## 2 PRELIMINARIES

In this section, we first provide a background on digital video data and then describe the Google Cloud Video Intelligence API.

### 2.1 Video Data

A digital video consists of audio data and a series of frames (still images) that are displayed in rapid succession to create the impression of movement. The frequency (rate) at which consecutive frames are displayed is called Frame Rate and is expressed in frames per second (fps). Modern video formats utilize a variety of frame rates. The universally accepted film frame rate is 24 fps. Some standards support 25 fps and some high definition cameras can record at 30, 50 or 60 fps [31].

Digital videos require a large amount of storage space and transmission bandwidth. To reduce the amount of data, video data are typically compressed using a lossy compression algorithm. Video compression algorithms usually reduce video data rates in two ways: 1) Spatial (intraframe) compression: Compressing individual frames, and 2) Temporal (interframe) compression: Compressing groups of frames together by eliminating redundant visual data across multiple consecutive frames, i.e., storing only what has changed from one frame to the next [31]. We refer to compression and decompression algorithms as *encoder* and *decoder*, respectively, and call the concatenation of encoder and decoder as *codec*.

### 2.2 Google Cloud Video Intelligence API

Success of ML algorithms has led to an explosion in demand. To further broaden and simplify the use of ML algorithms, cloud-based service providers such as Amazon, Google, Microsoft, BigML, and others have developed ML-as-a-service tools. Thus, users and companies can readily benefit from ML applications without having to train or host their own models.

Google has recently launched the Cloud Video Intelligence API for video analysis [25]. The API is designed to help better understand the overall content of the video, while providing temporal information on when each entity was present within the video. Therefore, it enables searching a video catalog the same way as text documents [30]. The API is made available to developers to deploy it in applications that require video searching, summarization or recommendation [30]. The API is said to use deep-learning models, built using frameworks like TensorFlow and applied on large-scale media platforms like YouTube [30].

A demonstration website has been also launched which allows anyone to select a video stored in Google Cloud Storage for annotation [25]. The API then provides the video labels (objects in the entire video), shot changes (scene changes within the video) and shot labels (description of the video events over time). As an illustration, Figure 2 shows the screenshots of API's outputs for the video "Animals.mp4," provided by the API website. Through experiments with different videos, we verified that the API's outputs are indeed highly accurate.

---
[1]The experiments are performed on the interface of Google Cloud Video Intelligence API's website in April 2017.

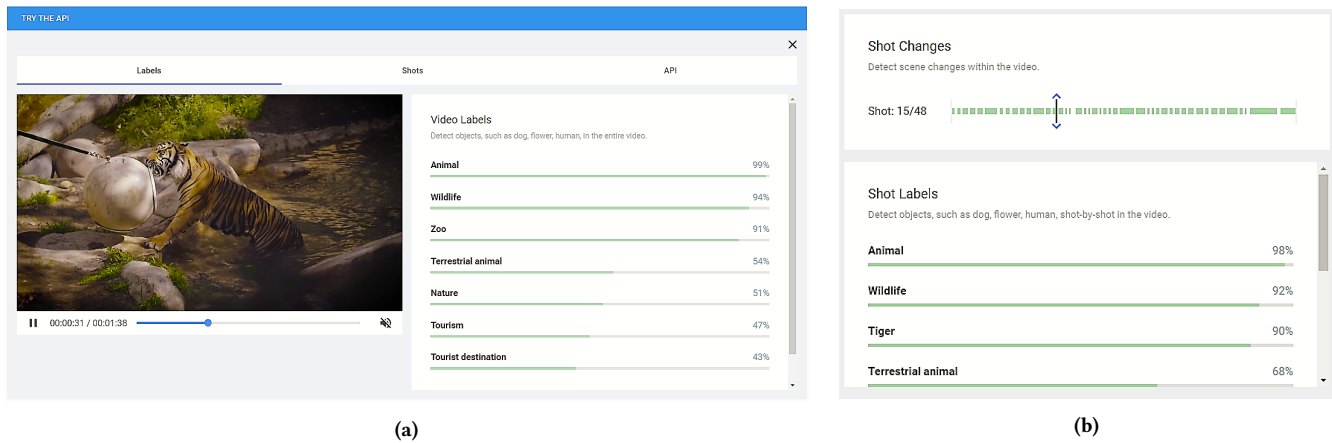

Figure 2: Screenshots of the API's outputs for "Animals.mp4" video, provided by the API website. a) Screenshot of the video labels, and b) screenshot of the shot changes and shot labels. The shot changes are the white bars appeared within the green strip. Shot labels are generated for each shot.

In our experiments, we queried and tested the API with different videos, including our recorded and synthetically generated videos, videos downloaded from web, and the sample videos provided by API website. Selected videos vary in content, length, frame rate, quality and compression format. For mounting the attacks, we modify videos, store them on Google cloud storage and then use them as inputs to the API. In one of our attacks, we insert images into the videos. Figure 3 shows some of the sample images that were used in our experiments. If not said otherwise, the manipulated videos are generated with frame rate of 25 fps, where each frame is a color image of size $300 \times 500$.

## 3 VIDEO ANALYSIS METHODS

In this section, we review the current methods for video classification and shot detection and provide a system model from the adversary's perspective for each task.

### 3.1 Video Classification Methods

Automatic video annotation would be a breakthrough technology, enabling a broad range of applications. It can help media companies with quickly summarizing and organizing large video catalogs. It can also improve video recommendations, as it enables the search engines to consider video content, beyond the video metadata. Another use case would be in video surveillance, where many hours of videos must be searched for a specific event. Moreover, Internet platforms, such as YouTube and Facebook, would be able to automatically identify and remove videos with illegal content.

Using ML techniques for video analysis is an active field of research [26, 27, 33, 57]. A simple approach is to treat video frames as still images and apply ML algorithms to recognize each frame and then average the predictions at the video level [58]. However, processing all video frames is computationally inefficient even for short video clips, since each video might contain thousands of frames. Moreover, consecutive video frames significantly overlap with each other in content and not all frames are consistent with the overall story of the video. Therefore, in order to learn a global description of the video while maintaining a low computational footprint, several papers proposed subsampling the video frames [18, 46, 50, 55, 58]. Hence, the focus of recent research is mostly on developing advanced ML techniques, such as deep neural networks, on subsampled frames. Figure 4a illustrates the model of video classification algorithms.

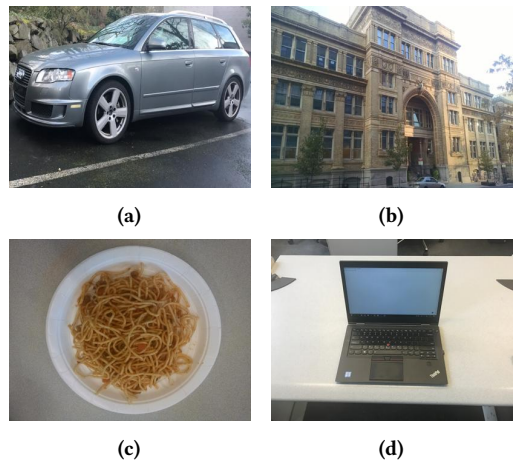

Figure 3: Sample images of (a) a car, (b) a building, (c) a food plate, and (d) a laptop, used in experiments.

### 3.2 Shot Detection Methods

Shot detection is used to split up a video into basic temporal units called shots, which is a series of interrelated consecutive pictures taken contiguously by a single camera and representing a continuous action in time and space. Shot detection is widely used in software for post-production of videos. It is also a fundamental step of automatic video annotation and summarization, and enables efficient access to large video archives [20]. In film editing, two methods are usually used to juxtapose adjacent shots: 1) Abrupt Transitions: A sudden transition from one shot to another, i.e. the last frame of one shot is followed by the first frame of the next shot, and 2) Gradual Transitions: The two shots are combined so that one shot is gradually replaced by another [20].

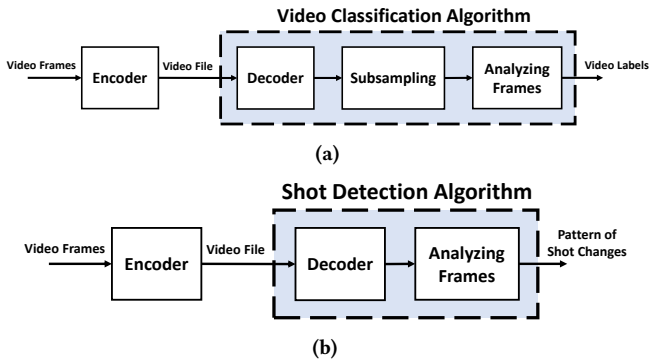

Figure 4: Models of video analysis algorithms. a) Video classification algorithm takes a video file, decodes it to obtain video frames, samples some of the frames and then processes subsampled frames to generate the video labels. b) Shot detection algorithm takes a video file, decodes it to obtain video frames and then processes the frames to generate the pattern of shot changes.

Shot detection algorithms try to find the positions in the video, where one scene is replaced by another one with different visual content. Conventional approaches for shot detection usually involve the two steps of measuring the similarity of consecutive frames and then determining the shots' boundaries. Two simple methods for measuring the similarity of frames are sum of absolute values of pixel-wise difference of frames and the difference between the histograms of frames. Compared to computing the pixel-wise difference of frames, the histogram-based method is more robust to minor changes within the scene. After scoring the difference of consecutive frames, typically a fixed or adaptive thresholding is used for localizing the shot changes [20]. Once a shot is detected, it can be treated as a short clip and the same algorithm for video classification can be applied for annotation. Figure 4b illustrates the model of shot detection algorithms.

## 4 THREAT MODEL

We assume that the video analysis system takes a video file and outputs video labels and the pattern of shot changes. We further assume that the system can only be accessed as a *black-box*. That is, the adversary possesses no knowledge about the training data or the specific ML models or video processing algorithms used, and can only query the system with any video of her choice and obtain the outputs. The goal of the adversary is to mount *targeted attacks*, i.e., given any video, the adversary intends to manipulate the video in such a way that the system generates only the adversary's desired outputs. The modification to the video should be very small, such that a human observer would perceive the content of the original, unmodified video. The proposed attacks are as follows:

- Targeted attack on video labeling algorithm: Deceiving the system to *only* output the adversary's desired video labels,
- Targeted attack on shot detection algorithm: Deceiving the system to output the adversary's desired pattern of shot changes.

The adversary's problem can be formulated as follows. Let $F$ be the video analysis algorithm and $\{X_t\}_{t=1:T}$ be the sequence of video frames, where $X_i$ is the $i$-th frame. For simplicity, we write $\{X_t\}_{t=1:T}$ as $X$. Let $y^*$ denote the adversary's desired output. Adversary's goal is to cause the algorithm to yield her desired outputs by making minimal changes to the video. Therefore, the adversary's objective function is as follows:

$$\text{Find } X^* \text{ s.t. } F(X^*) = y^* \text{ and } \|X - X^*\| \leq \epsilon,$$

where $X^*$ is the modified video. The term $\|X - X^*\|$ represents the amount of perturbation made to the video and $\epsilon$ is the maximum allowed perturbation, for which a human observer would still be able to perceive the content of the original video.

A video file can be modified in different ways. In the following, we review some methods of video modification and explain the maximum allowed perturbation corresponding to each method.

**Inserting images within video frames at a low rate:** The adversary can insert images of her choice within the video frames. However, the insertion rate must be low, so that the content of the original video would be perceivable. Moreover, it is preferable that the inserted image would be unnoticeable to a human observer. Speed of processing in the human visual system largely depends on the contrast of successive images shown [37, 48]. Empirical studies have shown that human visual system needs about 50 ms to process and comprehend every single image [17]. Therefore, in a video with frame rate of more than 20 fps, individual frames cannot be distinctly understood by humans. As a result, if an adversary inserts images at the rate of 1 fps within frames of such a video, the human observer would not perceive the content of inserted images.

**Removing video frames at a low rate:** Consecutive video frames are typically highly correlated, especially for videos with high frame rates, e.g., greater than 30 fps. Therefore, it is possible to remove video frames at a low rate (and replace them with adjacent frames) without significantly reducing the video quality. It is known that with frame rates greater than 20 fps, human visual system is fooled that the sequence of frames represents an animated scene, rather than being a succession of individual images [14]. Therefore, to preserve the video smoothness, removing video frames should not cause the rate of distinct video frames to drop below 20 fps.

**Slightly modifying individual frames:** Instead of adding or removing frames, the adversary can modify individual frames. The modification to each frame can be quantified by the PSNR value of the modified frame with respect to the original frame. For images $x$ and $x^*$ of size $d_1 \times d_2 \times 3$, PSNR value is computed as follows [52]:

$$PSNR = 10 \cdot \log_{10}\left(\frac{255^2}{\frac{1}{3 d_1 d_2} \sum_{i,j,k}(x_{i,j,k} - x^*_{i,j,k})^2}\right),$$

where $(i, j)$ is the image coordinate and $k \in \{1, 2, 3\}$ denotes the coordinate in color space. PSNR value is measured in dB.

Due to the inherent low-pass filtering characteristic of humans visual system, humans are capable of perceiving images slightly corrupted by noise [8], where acceptable PSNR values for noisy images are usually considered to be more than 20 dB [3]. However, similar to image insertion, an adversary is allowed to add high-density noise to video frames, but at a low rate, e.g., one frame per second. In contrast, the succession of noisy images corrupted even by low-density noise is very disturbing, especially when the video is passed through a lossy video codec. The reason is that video compression algorithms rely on similarity of consecutive frames to compress the video. A random uncorrelated noise added to frames increases the difference of consecutive frames. As a result, the compression algorithm reduces the quality of individual frames to achieve the same compression ratio.

In our attacks, we modify videos by low-rate image insertion within the frames or slightly perturbing frames, e.g., by adding low-density noise to frames at a low rate or smoothing some of the frames. The modifications are done in such a way that the content of the original video would be perceivable. In the following, we describe the attacks.

## 5 TARGETED ATTACK ON VIDEO CLASSIFICATION ALGORITHMS

In this section, we first present different approaches of attacking video classification algorithms and then describe our attack on Google Cloud Video Intelligence API.

### 5.1 Attack Approaches

For attacking a video classification algorithm, the adversary slightly manipulates the video. As shown in Figure 4a, the manipulated video passes through three blocks of codec, subsampling and frame processing. In the following, we discuss adversary's considerations for modifying videos.

For deceiving the system, the adversary can target the frame analysis algorithm. In this case, the adversary slightly modifies video frames such that, after subsampling, a human observer would classify the subsampled video as its original label, but the frame analysis algorithm yields adversary's desired labels. This attack type can be thought to be a form of generating adversarial examples [47] for frame analysis algorithm. However, as stated in Section 2.1, video files are typically compressed using lossy compression algorithms. The compression algorithm is likely to filter out the adversarial noise and thus render this approach ineffective. Therefore, the modification to video should be done in such a way that it would "survive" the codec operation. Moreover, even without the codec operation, this attack approach is challenging due to the adversary's black-box access to the model.

Another approach is to modify the video such that, after subsampling, a human observer would classify the subsampled video as adversary's desired label. This approach is preferable, since it is effective, regardless of frame analysis algorithm. For mounting this attack, the adversary inserts images with her desired content within the video frames, in locations where the subsampling function will sample. The subsampling locations need to be determined by querying the system with several specifically chosen videos. The success of inferring the subsampling function and the required number of queries depend on the function randomness and how it is related to video characteristics, such as video length, frame rate, codec, etc.

Table 1: Video labels generated by Google Cloud Video Intelligence API for different videos. The results suggest that the API's algorithm for generating video labels processes only the first frame of every second of the video.

| Video | Video labels returned by API |
|---|---|
| **Original Video:** A one-minute long video with all frames being the same image of a "building." | Building, Classical architecture, Neighbourhood, Facade, Plaza, Property, Apartment, Architecture, Mansion, Courthouse |
| **Modified Video 1:** Replacing the *first* frame of every second of the original video with an image of a "car." | Audi, Vehicle, Car, Motor vehicle, Land vehicle, Luxury vehicle, Sedan, Audi A6, Wheel, Mid-size car, Audi A4, Bumper, Family car, Audi RS 4 |
| **Modified Video 2:** Replacing the *second* frame of every second of the original video with an image of a "car." | Building, Classical architecture, Neighbourhood, Facade, Plaza, Property, Apartment, Courthouse, Mansion, Architecture |

We demonstrate the effectiveness of this approach by attacking the video and shot labeling algorithms of Google Cloud Video Intelligence API. Through experiments, we first infer the API's algorithm for sampling the video frames and then mount the image insertion attack. While the proposed attack is demonstrated on the Google API, this approach is applicable against any video classification system that is based on deterministic subsampling, e.g., [18, 46, 50, 55, 58] to name a few.

### 5.2 Inferring API's Algorithm for Sampling Video Frames

In order to infer the API's sampling algorithm, we first need to determine whether it uses deterministic or stochastic algorithms. For this, we queried the API with different videos. We found that when testing the API with the same video several times, it generates exactly the same outputs, in terms of the video and shot labels, the corresponding confidence scores, and the pattern of shot changes. Our observations imply that *the API's algorithms for processing the input videos are deterministic.*

Knowing that API's algorithms are deterministic, we designed an experiment for testing the API with videos which are different from each other on certain frames. We first generated a one-minute long video with all frames being the same image of a "building." We then modified the video in two different ways: 1) replacing the *first* frame of every second with an image of a "car," and 2) replacing the *second* frame of every second with the same image of the "car."

Table 1 provides the set of labels for the generated videos. As expected, all labels of the original video are related to "building," since it only contains images of a building. However, all labels of the first modified video are related to "car." In contrast, all labels of the second modified video are related to "building." We also tested the API with modified videos for which the $i$-th frame, $3 \leq i \leq f$, of every second is replaced by image of the "car," where $f$ is the number of frames per second. We observed that, for all of them, the API outputs video labels only related to "building." We selected videos with different characteristics and repeated the above experiments. We found that, regardless of the video content, length, frame rate, quality or compression format, *the API's algorithm for generating the video labels can be reduced to a function that gets as input only the first frame of every second of the video.* This suggests that the subsampling function samples video frames at the rate of 1 fps. Specifically, it samples the first frame of every second of the video.

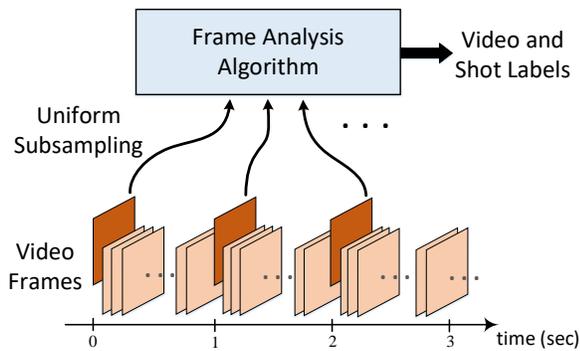

Figure 5: The video and shot labeling algorithm of Google Cloud Video Intelligence API subsamples the video frames uniformly and deterministically. Specifically, it samples the first frame of every second of the video.

Then, we examined how the API samples video frames for generating the *shot* labels. Similar to the case of video labels, we tested the API with a one-minute long video with all frames being the same image of a "building," and with modified videos, for which the $i$-th frame, $1 \leq i \leq f$, of every second is replaced by image of the "car." For the original video, there is only one shot and the shot labels are related to "building." As expected, the API generated 60 shots (one shot per second) for each of the modified videos, because each replaced frame is different from the adjacent frames and thus triggers a shot. For the first modified video, where image of the "car" is replaced in the first frame of every second, all shot labels were related to "car" for all 60 shots. In contrast, for $2 \leq i \leq f$, all the shot labels were related to "building." Our observations suggest that, similar to video labels, *the API generates shot labels by processing only the first frame of every second of the video.* Figure 5 illustrates the API's sampling algorithm.

### 5.3 Image Insertion Attack on API

Now, we describe the image insertion attack for changing the video and shot labels returned by the Google Cloud Video Intelligence API. The goal is to mount a targeted attack on the video classification algorithm. That is, given any video, the adversary intends to slightly manipulate the video, in such a way that a human observer would perceive the content of the original video, but the API outputs the adversary's desired video and shot labels. The attack procedure is as follows. The adversary is given a video and the target video label. She selects an image which represents the desired label, and inserts it into the first frames of every second of the video. The image insertion attack is illustrated in Figure 1.

Table 2: Results of Image Insertion Attack on Google Cloud Video Intelligence API. Sample videos are provided by API website [25]. The images are inserted in the first frame of every second of the video. The table only shows the video label with the highest confidence returned by API.

| Video Name | Inserted Image | Video Label Returned by API (Confidence Score) |
|---|---|---|
| "Animals" | "Car" | Audi (98%) |
| | "Building" | Building (97%) |
| | "Food Plate" | Pasta (99%) |
| | "Laptop" | Laptop (98%) |
| "GoogleFiber" | "Car" | Audi (98%) |
| | "Building" | Classical architecture (95%) |
| | "Food Plate" | Noodle (99%) |
| | "Laptop" | Laptop (98%) |
| "JaneGoodall" | "Car" | Audi (98%) |
| | "Building" | Classical architecture (95%) |
| | "Food Plate" | Pasta (99%) |
| | "Laptop" | Laptop (98%) |

For validating the attack on the API, we generated manipulated videos, stored them on Google cloud storage and used them as inputs to the API. Table 2 provides the API's output labels for the manipulated videos of three videos "Animals.mp4," "GoogleFiber.mp4" and "JaneGoodall.mp4," provided by API website (the table shows only the label with highest confidence score). As can be seen, regardless of the video content, the API returns a video label, with a very high confidence score, that exactly matches the corresponding inserted images. We tested the API with several videos with different characteristics and verified that the attack is consistently successful. We also mounted the image insertion attack for changing the shot labels returned by the API and verified that, by inserting an image in the first frame of every second of the video, *all* the shot labels returned by the API are related to the inserted image.

### 5.4 Improving Attack on API

In proposed attack, the image insertion rate is very low. For example, for a typical frame rate of 25, we insert only one image per 25 video frames, resulting in an image insertion rate of 0.04. Therefore, the modified video would contain the same content as the original video. Nevertheless, the attack can be further improved using the following methods.

**Lower insertion rate.** We found that the attack still succeeds if we insert the image at a lower rate of once every two seconds, i.e., by inserting the adversary's image into the video once every two seconds, the API outputs video labels that are only related to the inserted image. However, by further lowering the insertion rate, for some of the videos, the API's generated video labels were related to both the inserted image and also the content of the original video.

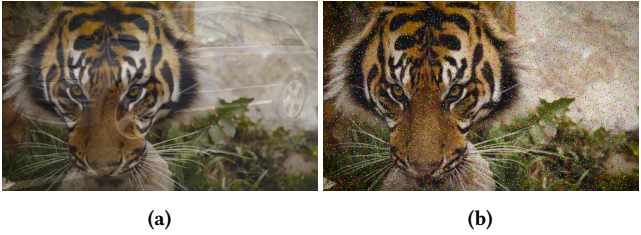

(a)          (b)

**Figure 6: Illustration of superposed and noisy images used for replacing the original video frames. a) The video frame is averaged with the image of a car, presented in Figure 3, where the weight of video frame is** $0.75$ **and the weight of the car image is** $0.25$**. b) The video frame corrupted by** $5\%$ **impulse noise. As can be seen, in both cases, the modification to the video frame is hardly noticeable.**

**Averaging with the original frame.** Instead of replacing the video frame, the adversary can *superpose* the image with the frame. Let $x_V$ be a video frame and $x_A$ be adversary's image. We generate $x'_V = \alpha x_V + (1-\alpha) x_A$ as the new frame and replace it for $x_V$. We found that by setting $\alpha = 0.75$, we can deceive the API to only output the video labels that are related to the inserted image. As an illustration, Figure 6a provides an example of a video frame averaged with the image of a car, presented in Figure 3. The weighted averaging of the adversary's image with the video frame significantly reduces the visual effect of the inserted image in the video, yet the API annotates the video as if it only contains the adversary's image. [2]

**Misclassification attack by adding noise.** In misclassification attack, the adversary's goal is to cause the API to output *different* video labels then the original labels. For mounting the misclassification attack, it is enough to just add noise to video frames. We performed the experiments with impulse noise. Impulse noise, also known as salt-and-pepper noise, is commonly modeled by [22]:

$$\tilde{x}_{i,j,k} = \begin{cases} 0 & \text{with probability } \frac{p}{2} \\ x_{i,j,k} & \text{with probability } 1-p \\ 255 & \text{with probability } \frac{p}{2} \end{cases}$$

where $x$, $\tilde{x}$ and $p$ are the original and noisy frames and the noise density, respectively. This model implies that each pixel is randomly modified to one of the two fixed extreme values, 0 or 255, with the same probability. Through experiments, we found that by adding only 5% impulse noise to the first frame of every second of the video, we can deceive the API to generate entirely irrelevant video labels. Figure 6b provides an example of a video frame corrupted by 5% impulse noise. As can be seen, the noise effect is hardly noticeable. We performed the experiments with other image noise types as well and found that impulse noise is the most effective one, i.e., it can cause the API to misclassify the input video by introducing very small perturbation to the video frames.

---

[2] We also tested the API with videos comprising of adversarial examples [47] and observed that the API is robust to this attack.

## 5.5 Applications of Image Insertion Attack

We showed that by inserting an image at the rate of 1 fps into the video, the video labels and all the shot labels generated by the API are about the inserted image. Instead of periodically inserting the same image, an adversary can replace the first frame of every second a video with frames from the corresponding positions of another video. The API then generates the same set of video labels for both videos, although they only have one frame in common in every second. In other words, the adversary can replace one frame per second of a video with the corresponding frame of another video and the API would not be able to distinguish the two videos.

Such vulnerability of video classification systems seriously undermines their applicability in real-world applications. For example, it is possible to poison a video catalog by slightly manipulating the videos. Also, one can upload a manipulated video that contains adversary's images related to a specific event, and a search engine wrongly suggests it to users who asked for videos about the event. Moreover, an adversary can bypass a video filtering system by inserting a benign image into a video with illegal content. Therefore, it is important to design video analysis algorithms to be robust and perform well in presence of an adversary. In the following, we provide a countermeasure against the image insertion attack.

## 5.6 Countermeasure Against Image Insertion Attack

One approach to mitigate the effect of the attack is to introduce randomness into the algorithms. In essence, inferring an algorithm that generates different outputs for the same input would be substantially more challenging, especially in the black-box setting. For video classification algorithms that subsample the video frames, an obvious countermeasure to the image insertion attack is to sample the frames randomly, while keeping the sampling rate the same.

Assume that the algorithm randomly samples the video frames at the rate of 1 fps and the adversary replaces $K$ video frames per second with her chosen image. Let the frame rate be $r$ fps. Thus, a fraction of $\frac{K}{r}$ sampled images are the inserted images by the adversary, which is equal to 4% for $r = 25$ and $K = 1$. Appendix A provides a detailed analysis on the probability that at least one adversary's image is sampled by the API, as well as the expected number of chosen adversary's images when the video frames are sampled at higher rates than 1 fps.

Analyzing the exact effect of sampling a small fraction of frames from adversary's images would require knowledge about the frame analysis algorithm and can be an interesting direction for future works. However, the process of the adversary uniformly replacing the video frames and the system randomly subsampling them is equivalent to the process of the adversary randomly replacing and the system uniformly sampling. Therefore, we can evaluate the performance of randomly subsampling the video frames using the current system, by randomly replacing the video frames with adversary's image. Through experiments, we found that even by randomly replacing two video frames every second, the API outputs mostly the same video labels as for the original video and none of the labels are related to the inserted image. The results imply that the video classification algorithm can benefit from randomly subsampling the video frames, without losing the performance.

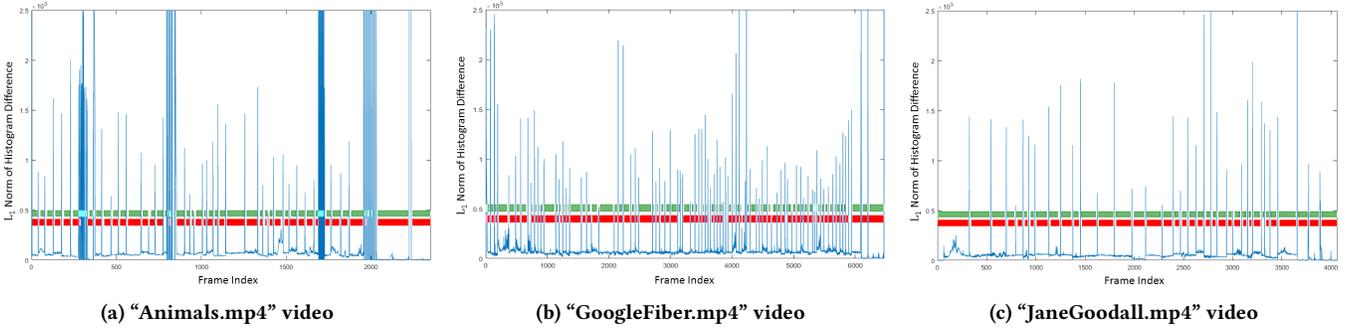

(a) "Animals.mp4" video    (b) "GoogleFiber.mp4" video    (c) "JaneGoodall.mp4" video

Figure 7: The histogram changes of consecutive video frames (blue curve), the shot changes generated by Google Cloud Video Intelligence API (green pattern) and our histogram-based method (red pattern). The shot changes returned by our method are located at large local maxima of the vector of histogram changes and are mostly aligned with API's output.

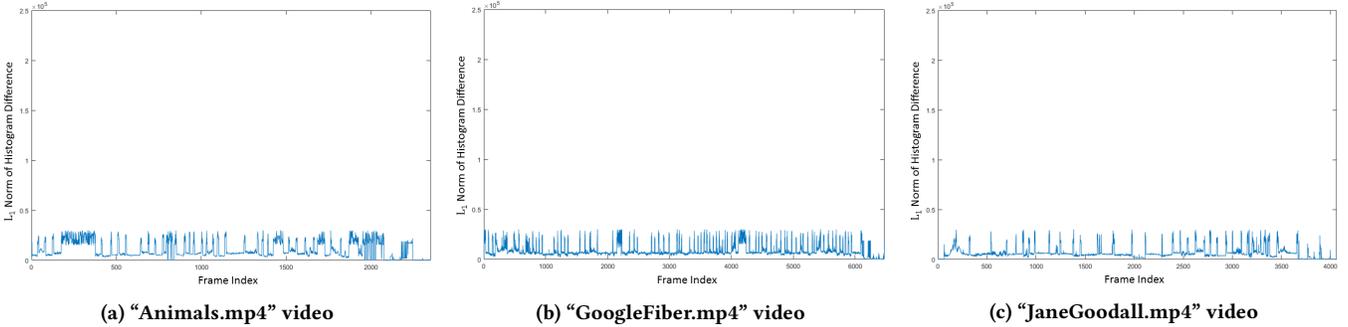

(a) "Animals.mp4" video    (b) "GoogleFiber.mp4" video    (c) "JaneGoodall.mp4" video

Figure 8: The histogram changes of consecutive frames for the smoothed videos. For the sake of comparison, we keep the y-axis of subfigures the same as the ones for Figure 7. All the peaks of the histogram changes are now smaller than our threshold of detecting a shot ($3 \times 10^4$). Therefore, our histogram-based method does not detect any shot change. We verified that the Google Cloud Video Intelligence API also generates only one shot for the entire smoothed video.

## 6 TARGETED ATTACK ON SHOT DETECTION ALGORITHMS

In this section, we first present our approach of attacking shot detection algorithms and then describe our attack on Google Cloud Video Intelligence API.

### 6.1 Attack Approach

For attacking a shot detection algorithm, the adversary needs to modify the video such that the algorithm detects a shot only at the adversary's desired locations. Since we assume that the adversary only has a black-box access to the system, for every video, she needs to query the system multiple times, each time with the video modified differently. This approach may require a large number of queries to the system, which can make the attack ineffective in real-world applications.

However, we noticed that different shot detection algorithms usually yield similar pattern of shot changes. Therefore, the adversary can rely on *transferability* of manipulated videos, i.e., a specific modification to a video causes different algorithms to detect the same pattern of shot changes. Hence, the adversary can first design her own shot detection algorithm and try to deceive it by manipulating the video. The adversary then uses the manipulated video as input to the target algorithm. We demonstrate the effectiveness of our approach by attacking the shot detection algorithm of Google Cloud Video Intelligence API. We first develop a histogram-based method for shot detection and then propose an attack for deceiving the algorithm to output our desired pattern of shot changes. We show that the manipulated videos can successfully deceive the API.

### 6.2 Histogram-Based Algorithm for Shot Detection

We first develop a method for shot detection and then compare the results with the ones generated by the API. Since shot changes are usually detectable on the gray-scale video, for simplicity, we transform the frames to gray-scale images. We adopt a *histogram-based algorithm*, where histogram of a frame is a vector that, for each gray-value, contains the number of pixels with that value. In this method, we first compute the difference between histograms of consecutive frames. If a frame has a large histogram difference with its previous frame, we declare that a shot is initiated at that frame. For measuring the difference of two histograms, we compute the $L_1$ norm of their difference, i.e., the sum of the absolute values of histogram changes.

Our method is described more formally in the following. Let $\{X_t\}_{t=1:T}$ be the set of video frames and $\{H_t\}_{t=1:T}$ be the set of histograms of gray-scale frames. We denote by $\delta H_t = H_t - H_{t-1}$ the element-wise difference of the histograms, which represents the change in statistics of the consecutive frames. Let $v$ be a vector of histogram changes of the video, obtained as the $L_1$ norm of the difference of histograms, i.e., $v_t = \|\delta H_t\|_1, 2 \leq t \leq T$. We locate the shot changes by finding the local maxima of the vector $v$, for which the amplitude of the histogram change is more than a threshold. The threshold is set to be greater than the typical histogram change of consecutive frames within a shot.

Through experiments with different videos, we explored whether the output of our method can resemble the pattern of shot changes returned by the API. Figure 7 provides the results for three sample videos, "Animals.mp4," "GoogleFiber.mp4" and "JaneGoodall.mp4." For each video, the figure shows the vector $v$ and the corresponding shot changes generated by the API (green pattern) and our histogram-based algorithm (red pattern). For our method, we set the threshold of $3 \times 10^4$ for declaring the shot change, which is determined by manually checking the histogram plots of several videos. We also discard very small shots (less than 10 frames). As can be seen, our method declares a shot change when there is a large local maximum in the vector of histogram changes. Also, the shot changes generated by our algorithm are mostly aligned with the API's generated pattern of shot changes. The results imply that *the API's algorithm of detecting shot changes can be reduced to an algorithm of finding the large peaks in the pattern of histogram changes of consecutive frames.*

### 6.3 Shot Altering Attacks on API

In this section, we present shot removal and shot generation attacks with the goal of deceiving the shot detection algorithm to miss real shot transitions and detect fake shots, respectively. Using the combination of shot removal and shot generation attacks, an adversary can deceive the API to output any desired pattern of shot changes for any video.

**Shot Removal Attack.** To force the API to miss a shot change, we need to modify an abrupt transition in the video to a gradual transition. For this, we apply a local low-pass filter on the frames located on the shot boundaries. We iteratively smooth those frames so that their histogram difference becomes less than the pre-specified threshold of our shot detection algorithm. Once we can deceive our histogram-based algorithm, we test the smoothed video on the API.

The smoothing attack is described more formally in the following. To remove the shot transition between the frames $X_t$ and $X_{t+1}$, we first smooth them as follows: $X'_t = \frac{1}{h}\sum_{i=-h}^{h} X_{t+i}$ and $X'_{t+1} = \frac{1}{h}\sum_{i=-h}^{h} X_{t+1+i}$, where $h = 1$. We then compute $v_t$, the $L_1$ norm of the difference of histograms of $X'_t$ and $X'_{t+1}$. If $v_t$ is greater than the threshold, we increase $h$ by one and repeat the process. We continue smoothing the frames, until the histogram change of the two frames becomes less than the threshold.

Through experiments with different videos, we examined whether the smoothed videos *transfer* to the API, i.e., the API also fails to detect the shot changes. We present the results on three sample videos, "Animals.mp4," "GoogleFiber.mp4" and "JaneGoodall.mp4," where the attack goal is to remove *all* shot changes, i.e., we want to deceive the API to believe the entire video is composed of only one shot. Figure 8 shows the histogram changes of the smoothed videos. For the sake of comparison, we keep the y-axis of subfigures the same as the ones for Figure 7. As can be seen, all the peaks now are smaller than the threshold of $3 \times 10^4$. Hence, our histogram-based method does not detect any shot change. We verified that the API also generates only one shot for the smoothed videos. Since our attack only smooths few frames on the shot boundaries and keeps the rest of the video the same, the perturbation to the video is hardly noticeable. Figure 9 illustrates an example of two frames on a shot boundary and their corresponding smooth frames. Unlike the original frames, the API does not detect any shot change between the smoothed frames.

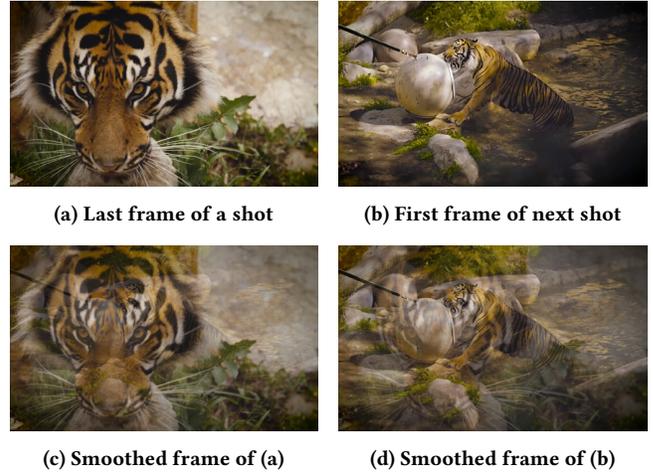

(a) Last frame of a shot  (b) First frame of next shot

(c) Smoothed frame of (a)  (d) Smoothed frame of (b)

**Figure 9: An Illustration of shot removal attack. Original frames represent an abrupt shot transition, while smoothed frames form a gradual shot transition.**

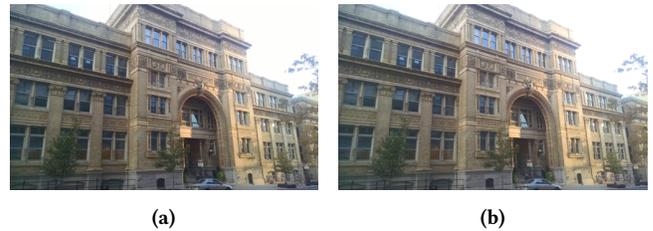

(a)  (b)

**Figure 10: An Illustration of shot generation attack. We generate a video, where all frames are an image of a "building" (subfigure (a)). We then select *one* of the video frames and increase all pixel values by $10$ (subfigure (b)). When testing the API with the modified video, it detects a shot change at the location of the modified frame.**

**Shot Generation Attack.** The goal of shot generation attack is to slightly modify the video such that the API wrongly detects a shot change at a specific frame. As explained in Section 5, inserting an image in between the video frames causes the API to declare a shot change at that frame. However, we do not need to insert an entirely different image within the frames to fake a shot change. We found that even by slightly increasing or decreasing all the pixel values of a frame, the API is deceived to detect a shot change at that frame.

As an example, we generated a one minute long video, where all frames were the same image of a building. As expected, the API outputs only one shot for this video. We then selected *one* of the video frames and increased all the pixel values by 10. When testing the API with the modified video, it detects a shot change at that frame. Figure 10 shows the original and modified frames. Notice that the frames are hardly distinguishable.

### 6.4 Applications of Shot Altering Attacks

Shot detection is important for temporal segmentation of the video and enables applications such as video searching and summarization. Fragility of the shot detection algorithm seriously undermines the benefits of the video analysis system. We showed that by slightly manipulating the video, an adversary can cause the API to generate her desired pattern of shot change. Shot altering attacks disrupt the functionality of the API by preventing it from correctly indexing and annotating the video content. Essentially, without correctly detecting the individual shots, the system cannot reliably produce the story of the video by cascading the proper shot labels. As an example of the attack, an adversary can prevent the API from finding important events in video surveillance data. In the following, we provide a countermeasure against shot altering attacks.

### 6.5 Countermeasure Against Shot Altering Attacks

Several papers have proposed methods for gradual shot detection in videos [5, 20, 45]. We are however interested in robust techniques for *adversarial* inputs, i.e., videos that are maliciously modified to deceive the algorithm. Similar to the image insertion attack, the system robustness can be improved by introducing randomness to the algorithm, e.g., random subsampling of the video frames and measuring the difference of sampled frames as an indicator for whether the shot has changed. However, compared to the task of video labeling, the system needs to subsample video frames at higher rates, e.g., 10 fps, to be able to detect small shots. Also, since the scene may change within the shot, samples within the shot may have large differences. Therefore, compared to consecutive frames, a higher threshold needs to be set when measuring the difference of subsampled frames. Moreover, the accuracy can be improved by running the randomized algorithm several times and detecting the shot changes by averaging the results.

There is a trade-off between accuracy and robustness of the shot detection algorithm. Sampling at lower rates increases the robustness to spurious changes within the shots, but causes the algorithm to potentially miss the small shots occurred between the two consecutive sampled frames. Moreover, with subsampling, the information of the exact moment of shot transition will be lost; but, since the intersampling times are less than one second, the estimated time suffices for most applications. Random subsampling, however, will not affect the shot labels; therefore, the overall description of the video remains mostly the same.

## 7 RELATED WORK

With the proliferation of mobile devices, video is becoming a new way of communication between Internet users. Accelerated by the increase in communication speed and storage space, video data has been generated, published and spread explosively. This has encouraged the development of advanced techniques for various applications, including video search and summarization [55]. However, due to the temporal aspect of video data and the difficulty of collecting sufficient well-tagged training samples [51, 60], using machine learning for videos classification has remained a challenging task. In recent years, several new datasets have been published to help advancing the field [13, 27]. Most recently, a new large dataset [2] and a competition [16] is announced by Google to further promote the research in video classification methods.

Following the successful use of deep neural networks for image classification [21, 28, 44], researchers have attempted to apply deep learning techniques to the video domain. The current research for video analysis has mainly focused on selecting the network architectures and also the inputs to the networks. From network architecture perspective, convolutional neural networks [26, 57], recurrent neural networks [58], and temporal feature pooling [27, 58] are viewed as promising architectures to leverage spatial and temporal information in videos.

For network inputs, current works usually adopt subsampling the video frames, in order to gain both computational efficiency and classification accuracy [18, 46, 50, 55, 58]. To compensate for the loss of motion information, [43] suggested incorporating explicit motion information in the form of optical flow images computed over adjacent frames. Optical flow encodes the pattern of apparent motion of objects and is computed from two adjacent frames sampled at higher rates than the video frames [59]. It has been however observed that the additional gain obtained by incorporating the optical flow images is very small [58]. Also, in [54], the authors proposed using audio spectrogram along with the visual data to improve the video understanding.

The security of ML models has been studied from different perspectives [4, 6, 7, 10, 36]. Learning models are subject to attacks at both training and test phases. The main threat during training is poisoning attack, which is injecting fake samples into the training data or perturbing the training samples in order to influence the model [11, 24]. During test time, the adversary may try extract private training data, e.g., medical data, from the trained model [42]. To protect the training data, several papers have proposed algorithms for privacy preserving ML models [1, 41]. Another threat is extracting model parameters by querying the model multiple times [49]. It has been also shown that ML models can be deceived by perturbing the input features or generating out-of-distribution samples [9, 19, 29, 32, 35, 47, 56]. This property has been used to attack voice interfaces [15], face-recognition systems [40] and text classification systems [23, 39].

Attacks on ML systems can be also classified according to the adversary's access to the system. In white-box model, the adversary is assumed to have some information about the learning algorithm, while the black-box model assumes that the adversary only has an oracle access to the system [36]. One attack approach in black-box scenario is to generate adversarial data based on a substitute model and transfer them to the target model [34].

In this paper, we developed targeted attacks on video analysis algorithms, by slightly perturbing the input videos, and applied the attacks on a real-world video analysis system. The attacks are developed with having only a black-box access to the system. Moreover, unlike the existing black-box attacks on ML models [34], we had no information about the training data or even the set of output labels of the model. We suggested that introducing randomness to the video analysis algorithms can improve the system robustness.

## 8 CONCLUSION

In this paper, we proposed attacks on two fundamental classes of automatic video analysis algorithms, namely video classification and shot detection. We then applied the attacks on Google Cloud Video Intelligence API. We showed that the system can be easily deceived by an adversary without compromising the system or having any knowledge about the specific details of the algorithms used. By conducting several experiments, we inferred that the API generates video and shot labels by sampling the video frames deterministically and uniformly. We showed that by replacing an image into the first frame of every second the video, the API's generated video and shot labels are only related to the inserted image. We also showed that the API's output pattern of shot changes can be mostly recovered by finding the large peaks of the vector of histogram changes. Based on this observation, we proposed two shot altering attacks for generating fake shot changes and removing the real ones. We suggested that using stochastic algorithms can improve the robustness of the API and proposed random subsampling as a countermeasure for the proposed attacks.

As another observation, we found that, in line with most of the state-of-the-art methods, the API does not use the audio data for video annotation. Specifically, we observed that the API generates the same outputs for videos with the same visual data, but different audio data. As a result, an attacker can embed illegal or inappropriate audio data in a benign video and evade the video screening filters. With ML systems being deployed in daily applications, the success of our attacks further indicates the importance of designing learning systems to be robust in real-world applications.

## APPENDIX
## A ANALYSIS OF RANDOM SUBSAMPLING

In this section, we analyze the random subsampling method. Let the frame rate be $r$ fps. Suppose that the adversary uniformly (and without replacement) selects a set of $K$ out of the $r$ frames to replace. Suppose further that the system chooses $L$ samples out of the $r$ frames, uniformly at random and without replacement for analysis. We compute the probability that at least one of the adversary's images is chosen by the system, as well as the expected number of sampled adversary's images.

**Probability of sampling an adversary's image.** Let $Y$ denote a random variable representing the number of images that are sampled from the adversary's images. We are interested in computing $Pr(Y \geq 1)$, or equivalently $1 - Pr(Y = 0)$. Let $\chi_i$ denote the event that the $i$-th sample is from original video frames. We have

$$
\begin{aligned}
1 - Pr(Y = 0) &= 1 - Pr(\chi_1 \cap \cdots \cap \chi_L) \\
&= 1 - \prod_{i=1}^{L} Pr(\chi_i | \chi_1, \ldots, \chi_{i-1}) \quad (1) \\
&= 1 - \prod_{i=1}^{L} \left( \frac{r - K - (i-1)}{r - (i-1)} \right) \quad (2)
\end{aligned}
$$

where (1) follows from the conditioning on the $\chi_i$'s and (2) holds since the probability of choosing a valid sample, given that the first $(i-1)$ samples are valid, is equal to the number of remaining valid samples $(r - L - (i-1))$ divided by the total number of remaining samples $r - (i-1)$. For $K, L \ll r$, this can be approximated by $Pr(Y \geq 1) \approx 1 - e^{\frac{-KL}{r}}$, which can be further approximated by $Pr(Y \geq 1) \approx \frac{KL}{r}$, when $KL \ll r$. Also, for $L = 1$, we have $Pr(Y = 1) = \frac{K}{r}$, meaning that the chance that adversary's image will be sampled increases linearly by the number of replaced images.

**Expected number of selected images from adversary.**
Let $T(L, K, r) \triangleq \mathbf{E}(Y)$ denote the expected number of adversary's images that are chosen, and let $\chi_1$ be defined as above. Let $\overline{\chi}_1$ denote the complement of $\chi_1$. We have that

$$
\begin{aligned}
T(L, K, r) &= \mathbf{E}(X | \chi_1) Pr(\chi_1) + \mathbf{E}(X | \overline{\chi}_1) Pr(\overline{\chi}_1) \\
&= T(L-1, K, r-1)(1 - \frac{K}{r}) + T(L-1, K-1, r-1)\frac{K}{r}.
\end{aligned}
$$

where $T(L, K, K) = \min\{L, K\}$. Hence, the expected number of adversarial samples can be computed recursively. For $K, L \ll r$, we have $\mathbf{E}(Y) \approx \frac{KL}{r}$. With larger $L$, the number of frames sampled from adversary's image increases, however the ratio of adversarial frames to all sampled frames remains approximately the same.

## ACKNOWLEDGMENTS

This work was supported by ONR grants N00014-14-1-0029 and N00014-16-1-2710, ARO grant W911NF-16-1-0485 and NSF grant CNS-1446866.